\documentclass[floatfix,10pt,onecolumn,showpacs,amsmath,amssymb]{revtex4}

\usepackage{epsf}
\usepackage{graphicx}  
\usepackage{dcolumn}   
\usepackage{bm}        

\newcommand{\be}{\begin{equation}}
\newcommand{\en}{\end{equation}}
 \newcommand{\bea}{\begin{eqnarray}}
 \newcommand{\ena}{\end{eqnarray}}
  \newcommand{\sch}{Schwarzschild}

\begin{document}

\title{Misner-Sharp Mass in $N$-dimensional $f(R)$ Gravity}
\author{Hongsheng Zhang$^{1,3~}$\footnote{Electronic address: hongsheng@shnu.edu.cn}, Yapeng Hu$^{2,3}$, Xin-Zhou Li$^1$ \footnote{Electronic address: kychz@shnu.edu.cn} }
\affiliation{ $^1$ Center for
Astrophysics, Shanghai Normal University, 100 Guilin Road,
Shanghai 200234, China\\
$^2$ College of Science, Nanjing University of Aeronautics and Astronautics, Nanjing 210016, China\\
$^3$ State Key Laboratory of Theoretical Physics, Institute of Theoretical Physics, Chinese Academy of Sciences, Beijing, 100190, China
}

\date{ \today}

\begin{abstract}
We study the Misner-Sharp mass for the $f(R)$ gravity in an $n$-dimensional (n$\geq$3) spacetime which permits three-type  $(n-2)$-dimensional maximally symmetric subspace. We obtain the Misner-Sharp mass via two approaches. One is the inverse unified first law method, and the other is the conserved charge method by using a generalized Kodama vector. In the first approach, we assume the unified first still holds in the $n$-dimensional $f(R)$ gravity, which requires a quasi-local mass form (We define it as the generalized Misner-Sharp mass). In the second approach, the conserved charge corresponding to the generalized local Kodama vector is the generalized Misner-Sharp mass. The two approaches are equivalent, which are bridged by a constraint. This constraint determines the existence of a well-defined Misner-Sharp mass. As an important special case, we present the explicit form for the static space, and we calculate the Misner-Sharp mass for Clifton-Barrow solution as an example.

\end{abstract}

\pacs{04.20.-q, 04.70.-s}
\keywords{Misner-Sharp mass; f(R) gravity; conserved charge}

\preprint{arXiv: }
 \maketitle

\section{Introduction}

  In opposite to intuition, the mass (energy) of gravity field is an extraordinarily involved problem. The existence of tensor form of gravitational stress-energy is forbidden by equivalence principle. At least we can define the total mass at spacelike infinity (Arnowitt-Deser-Misner mass) and null infinity (Bondi-Sachs mass) for asymptotic flat manifolds. This situation cannot satisfy us because we can consider neither the gravity energy flux between different regions nor the energy flux between gravity field and other fields. It is thus hard to investigate the relativistical astrophysics processes. In view of this situation, physicists introduced the notion of quasilocal mass. Quasilocal mass is defined on a finite compact two-surface. Now we have several different quasilocal mass forms, for instance Penrose mass, Hawking-Hayward mass, Chen-Nester mass, Brown-York mass, etc. For the motivations and general rules they should obey, see the review article \cite{qloc}.

  In a spherically symmetric spacetime, one of the leading mass form is the Misner-Sharp mass. Misner and Sharp defined it in the studies of gravitational collapse \cite{ms}. Subsequent works show that it has several nice properties \cite{hayward0, hayward1}. In the Newtonian limit, it reduces to the Newtonian mass to the leading order and Newtonian mass plus Newtonian potential energy to the next to leading order. It is the conserved charge corresponding to the timelike Killing vector in static spherically symmetric spacetime and to the Kodama vector in a general spherically symmetric spacetime. Also, it is useful when quantum effect is included \cite{nil}. The Misner-Sharp mass is just the~\sch~mass in the ~\sch~ space, which is independent of radius. We find that Misner-Sharp mass is the energy in an adiabatic system. Under this assumption, we can derive ~\sch~ solution by the first law, without invoking Einstein field equation \cite{self}.

  Recently, the Misner-Sharp mass is extended to 4 dimensional $f(R)$ gravity \cite{cai}. The $f(R)$ gravity is a kind of higher derivative gravity theory. The Hilbert-Einstein action plus a cosmological constant term is the unique Lagrangian which can generate field equations without higher than two order derivatives with respect to metric in 4 dimensional spacetime. However, the studies of renormalizability of gravity imply that we indeed need higher order derivatives in the field equation to control divergence \cite{stelle}. Moreover, the higher order derivatives always appear in the low energy effective field equation when we consider the quantum corrections \cite{birrell}, or string theory \cite{string}. Such motivations urge people to consider the higher derivative gravity seriously. $f(R)$ theory is one of the most thoroughly studied model in higher derivative theories. In this model, the Lagrangian is an arbitrary (smooth) function of the Ricci scalar $R$. Compared to more generic and more complicated models including terms like $R_{\mu\nu}R^{\mu\nu}$, $R_{\alpha\beta\gamma\delta}R^{\alpha\beta\gamma\delta}$, it has at least two distinct advantages: First of all, it is simple. The Lagrangian $f(R)$ is general enough to encode the fundamental structure of higher derivative theories, and simple enough to handle. Second, it is the unique one which can evade the critical Ostrogradski instability among all higher derivative gravity theories \cite{wood}. For a review of $f(R)$ model, see for example \cite{soti}.

  Higher dimensional theory has a fairly long history since Kluza-Klein's proposal.  Higher dimension is even treated as a universal feature of unified theory including gravity. Similar to the 4-dimensional case, we need higher order derivatives to control the divergence. Note that the combinations of Riemann tensor, Ricci tensor and Ricci scalar which do not generate higher order derivatives in field equations, called Lovelock term, is also widely investigated in higher dimensional theories. On the other hand the lower dimensional gravity theory is also of interests. 3-dimensional gravity is very useful to study the 2-dimensional conformal field theory, and is relatively easy to get analytical results, for a review see \cite{3dgra} and reference therein. Recently, we find a black hole and gravity wave solutions in the frame of 3-dimensional $f(R)$ gravity \cite{zll}. In this article, we shall explore the Misner-Sharp mass for $n$-dimensional $f(R)$ gravity ($n\geq 3$).

    This paper is organized as follows. In section 2 we review some previous results of Misner-Sharp mass, stressing on the inverse unified first law method \cite{cai}. In section 3 we obtain the Misner-Sharp mass for $f(R)$ gravity in $n$-dimensional spacetime ($n\geq 3$) by inverse unified first law method and conserved charge method, respectively. In section 4, we present the special result in a static spacetime. In section 5, we conclude this article.

    \section{Misner-Sharp mass: some previous results}

   In 4 dimensional spherically symmetric spacetime, we adopt the following coordinates,
   \be
   ds^2=I_{ab}dx^adx^b+r^2d\Omega_2^2,
   \label{2dim}
   \en
   where $\Omega_2$ denotes a unit 2-sphere, $I_{ab}$ are general functions which are independent on the inner coordinates of the 2-sphere, and $a,~b$ run from 0 to 1. This metric can be dynamic or static. In this spacetime, the Misner-Sharp mass can be defined as,
   \be
   M_{ms}=\frac{r}{2G}\left(1-I^{ab}\partial_a r\partial_b r\right).
   \label{defi}
   \en
  With this definition of Misner-Sharp mass, we have the unified first law \cite{hayward1},
  \be
  dE = A \Psi_a dx^a + W dV,
  \label{uni}
\en
   where $V=4\pi r^3/3$ is the volume of the space region in consideration, $A=4\pi r^2$ is its surface area. $W$ is the work term,
     \be
     W=-\frac{1}{2}I^{ab}T_{ab}.
     \en
 Here $T_{ab}$ denotes the stress-energy tensor of the matter field. $\Psi_a$ labels the
 energy supply term,
 \be
 \Psi_a=T_a^{\ b} \partial _b r +W \partial_a r.
 \en
  The unified first law has been used in several different situations of Einstein gravity and in modified gravity \cite{cai1, lihui}.
  The logic in the unified first law is as follows: we first define the Misner-Sharp mass as shown in (\ref{defi}), and then we find that it satisfies the unified first law (\ref{uni}). However, in the modified gravity theories, we have no prior definition of Misner-Sharp mass. Exactly the unified first law requires a mass in the LHS of (\ref{uni}). Thus we can define the mass in (\ref{uni}) as an extension of Misner-Sharp mass in modified gravity theories by using the RHS of (\ref{uni}) \cite{cai}.

  Now we demonstrate the inverse first law method for a concrete metric in modified gravity.  We work in an $n$-dimensional spacetime. Without losing generality, we write an $n$-dimensional spacetime with $(n-2)$-dimensional maximally symmetric submanifold in a double-null coordinates
  \be
ds^{2}=-2e^{-\varphi(u,v)}dudv+r^{2}(u,v)\gamma _{ij}dz^{i}dz^{j}.
\label{metricn}
\en
 Here $\gamma _{ij}$ dentoes the metric of the $n-2$ dimensional
 maximally symmetric submanifold with sectional curvature
$k=-1, 0, +1$. The whole manifold $\cal M$ is a direct product of this maximally $(n-2)$-dimensional manifold $({\cal K}, \gamma)$ and a 2 dimensional manifold $({\cal T}, I)$, where $I_{uv}=I_{vu}=-e^{-\varphi(u,v)}$.

Similar to
the case of Einstein gravity (\ref{uni}), we suppose that the
 unified first law takes the same form
\begin{equation}
dM_{ms}=A\Psi_a dx^a +WdV,
 \label{uni2}
\end{equation}
 but $A$ and $V$ correspond to the region of the $n-2$ dimensional maximally symmetric submanifold with radius $r$, i.e., $A=V_{n-2}^{k}r^{n-2}$ and $V =V_{n-2}^{k}r^{n-1}/(n-1)$. The energy supply  term $\Psi_a $ and work density $W$ are defined on $({\cal T}, I)$. The concrete $\Psi_a$ and $W$ take the same forms of Einstein gravity.
In the coordinates (\ref{metricn}) the RHS in (\ref{uni2}) reads,
\be
A\Psi_a dx^a +WdV=A(u,v)du+B(u,v)dv,
 \label{AB}
\en
where
\be
\label{AB1}
A(u,v)=V_{n-2}^{k}r^{n-2}e^{\varphi}(r,_{u}T_{uv}-r,_{v}T_{uu}),
\en
\be
B(u,v)
=V_{n-2}^{k}r^{n-2}e^{\varphi}(r,_{v}T_{uv}-r,_{u}T_{vv}).
\label{AB2}
\en
 Here a comma denotes partial derivative. Substituting (\ref{uni2}) into (\ref{AB}), we reach
 \be
 F\equiv dM_{ms}=A(u,v)du+B(u,v)dv.
 \label{F}
 \en
 The above expression is only a formal equality. $M_{ms}$ is well-defined only if $F$ is a locally exact form. And thus $F$ is a closed form, from which we can obtain the constraint for the existence of a well-defined $M_{ms}$. The essence of the above discussions is that we assume the unified first law takes the same form as in 4-dimensional Einstein gravity, and then we derive the required mass in the unified first law. This is different from the Einstein case where we first define Misner-Sharp mass and then prove it satisfies the unified first law. So we call this method ``inverse unified first law'' method.

      \section{Misner-Sharp mass for $n$-dimensional $f(R)$ gravity}
       In this section we explore the Misner-Sharp mass for $f(R)$ gravity in an $n$-dimensional spacetime via unified first law method and conserved charge method, respectively. And we shall demonstrate the two methods are equivalent.

       The action of the $f(R)$ gravity in an $n$-dimensional spacetime reads
   \be
   S=\frac{1}{16\pi G }\left(\int_{\cal M} d^{n}x\sqrt{-{\rm det}(g)}~f(R)+\int_{\cal \partial M} d^{n-1}x\sqrt{-{\rm det}(h)}~2B_o\right)+S_{m}.
  \label{action}
  \en
 Here $G$ denotes the $n$-dimensional Newtonian constant, $B_o$ represents the corresponding boundary term for $f(R)$ term, $g$ is the $n$-dimensional metric, $h$ denotes the induced metric on the boundary, and $S_m$ labels the action of matter fields. The field equation follows the action (\ref{action}) can be written as,
\be
f_{R}R_{\mu \nu }-\frac{1}{2}fg_{\mu \nu }-\nabla _{\mu }\nabla
_{\nu }f_{R}+g_{\mu \nu }\square f_{R}=8\pi G T_{\mu \nu },
\label{field}
\en
where $T_{\mu \nu }$\ denotes the stress-energy for matter fields, and $f_R=\frac{\partial f}{\partial R}$. We work on the manifold $({\cal M},~g)$ defined in (\ref{metricn}). The components of the field equation (\ref{field}) in the coordinates (\ref{metricn}) read,

\bea
-8\pi GT_{uu}&=& (n-2)f_{R}\frac{\varphi,_{u}r,_{u}+r,_{uu}}{r}+f_{R},_{uu}
+\varphi,_{u}f_{R},_{u},  \notag \\
-8\pi G T_{vv}&=& (n-2)f_{R}\frac{\varphi,_{v}r,_{v}+r,_{vv}}{r}+f_{R},_{vv}
+\varphi,_{v}f_{R},_{v},  \notag \\
8\pi G T_{uv}&=& f_{R}\varphi_{,uv}-(n-2)f_{R}\frac{r,_{uv}}{r}+\frac{1}{2%
}fe^{-\varphi}+f_{R},_{uv}+\frac{n-2}{r}({r,_{u}f_{R},_{v}+r,_{v}f_{R},_{u}}).
\label{Tcompt}
\ena

    Clearly, the components of the stress-energy tensor inherit the symmetry between u and v in the metric. We note that the sectional curvature of the submanifold $\cal K$ does not appear in these components, as shown above. This property is a result of $\cal M$=${\cal T}\times {\cal K}$. $k$ shows itself in the Ricci scalar $R$ of the whole manifold $\cal M$,
    \be
    R=\frac{k(n-3)(n-2)}{r}+2e^\varphi\left(\frac{(n-3)(n-2)r_{,u}r_{,v}}{r^2}-\varphi_{,uv}+\frac{2(n-2)r_{,uv}}{r}\right).
    \label{RS}
    \en
    A well-defined $M_{ms}$ in (\ref{F}) requires $F$ is a closed form $dF=0$, which means,
    \be
    A_{,v}~ dv\wedge du+B_{,u}~ du\wedge dv=0.
    \en
    Then we obtain the constraint for a well-defined $M_{ms}$,
     \be
     A_{,v}=B_{,u}.
     \label{AvBu}
     \en
    Substituting (\ref{AB1}) and (\ref{AB2}) into (\ref{AvBu}), where the stress-energy is shown in (\ref{Tcompt}), we obtain the constraint,
    \be
    (\varphi,_{u}r,_{u}+
 r,_{uu})(f_{R},_{vv}+\varphi,_{v}f_{R},_{v})
  =(\varphi,_{v}r,_{v}+r,_{vv})(f_{R},_{uu}+\varphi,_{u}f_{R},_{u}),
  \label{condi}
    \en
    which is independent on dimension of the spacetime and the sectional curvature of $\cal K$. The
   derivation of the above equation is rather complicated, where we have used
   \be
   f_{,u}=f_RR_{,u},
   \en
   \be
   f_{,v}=f_RR_{,v},
   \en
    and $R$ is given by (\ref{RS}).
   Under this constraint, $F$ is closed and thus $M_{ms}$ is well-defined. $M_{ms}$ can be calculated by
   \be
   M_{ms}=\int Bdv,
   \en
   or
   \be
   M_{ms}=\int Adu.
   \en
  One is easy to show that the above two equations are equivalent under the constraint (\ref{AvBu}). The latter one yields,
  \bea
  M_{ms}=\frac{V_{n-2}^{k}r^{n-3}}{8\pi G }\Big[\frac{n-2}{2}(k-I^{ab}\partial _{a}r\partial_{b}r)f_{R}+\frac{1}{2(n-1)}%
r^{2}(f-f_{R}R)- rI^{ab}\partial _{a}f_{R}\partial _{b}r\Big] ~~~~~~~~~~~~~~~~~~~~~~~~~~\notag \\
-\frac{V_{n-2}^{k}}{8\pi G }\int du\Big[f_{R},_{u}e^{\varphi}(r^{n-2}r,_{v}),_{u}+f_{R},_{u}\Big(\frac{(n-2)k}{2}r^{n-3}-%
\frac{1}{2(n-1)}r^{n-1}R\Big)+f_{R},_{v}r^{n-2}(r,_{u} e^{\varphi}),_{u}\Big].
\label{MS}
  \ena
  Thus we complete the generalization of Misner-Sharp mass to higher dimension case for $f(R)$ gravity by using one of its most important property: It satisfies the unified first law. Then there comes a question: Is this a reasonable generalization of Misner-Sharp mass in the $n$-dimensional $f(R)$ gravity? To find the answer is an arduous task, but this generalization can be reintroduced by another way.  The other important property of Misner-Sharp mass is that it is associated with the conserved charge corresponding Kodama vector in a spherically symmetric spacetime. One can extend the Misner-Sharp mass to Gauss-Bonnet gravity by using this property \cite{mae}, which is equivalent to the inverse first law method \cite{cai}. Here we check whether the two methods are equivalent in the $n$-dimensional $f(R)$ gravity case.

  In Ref. \cite{koda}, Kodama vector is defined in a spherically symmetric spacetime. Now we generalize the concept of Kodama vector for the spherically symmetric case to one with maximally symmetric submanifold. The generalized Kodama vector is defined as,
\be
K^{\mu }=-\epsilon ^{\mu \nu }\nabla _{\nu}r,
\label{kodama}
\en
 on $({\cal M},~g)$, where $\epsilon _{\mu \nu }=\epsilon _{ab}(dx^{a})_{\mu }(dx^{b})_{\nu}$, and $
\epsilon _{ab}$ is the compatible volume element to the metric $I$ on ${\cal T}$. The Greek indexes $\mu,~\nu$ run from $0$ to $n-1$.  This is a straightforward generalization of Kodama tensor. From direct calculation, (\ref{kadama}) can be rewritten as,
\be
K^{\mu}=e^{\varphi}\left[r,_{v}\Big(\frac{\partial }{\partial
u}\Big)^{\mu} -r,_{u}\Big(\frac{\partial }{\partial
v}\Big)^{\mu}\right].
\label{kodama2}
\en
 Just like the case of a Killing vector, the Kodama vector induces a conserved current in Einstein gravity. Mimicking the case of Einstein gravity, we define
 \be
 J^{\mu }=-T^{\mu }{~}_{\nu}K^{\nu}.
  \label{Jmu}
 \en
 However, we must check the conservativeness of $J^{\nu}$ in the $n$-dimensional $f(R)$ gravity. Firstly, Bianchi identity ensures that the stress-energy of matter field $T_{\mu\nu}$ is still conserved,
 \be
 \nabla _{\mu}T^{\mu\nu}=0.
 \en
 Then
  $J^{\nu}$ is conserved,
  \be
   \nabla _{\mu}J^{\mu}=0,
   \en
   only if,
   \be
   \nabla _{\mu }\nabla _{\nu}(f_{R}\nabla ^{\mu }K^{\nu})=0.
   \label{con}
   \en
   With the aid of $J$, we define a charge,
   \be
   Q_J=\int_{S} {}^*J~,
   \en
   where a star denotes the Hodge dual of the vector $J$. $S$ is an $n-1$-dimensional hypersurface in $\cal M$. $Q_J$ is really a conserved charge under the constraint (\ref{con}). Slicing $\cal M$ with constant $v$, we reach,
    \bea
  Q_J=\frac{V_{n-2}^{k}r^{n-3}}{8\pi G }\Big[\frac{n-2}{2}(k-I^{ab}\partial _{a}r\partial_{b}r)f_{R}+\frac{1}{2(n-1)}%
r^{2}(f-f_{R}R)- rI^{ab}\partial _{a}f_{R}\partial _{b}r\Big] ~~~~~~~~~~~~~~~~~~~~~~~~~~\notag \\
-\frac{V_{n-2}^{k}}{8\pi G }\int du\Big[f_{R},_{u}e^{\varphi}(r^{n-2}r,_{v}),_{u}+f_{R},_{u}\Big(\frac{(n-2)k}{2}r^{n-3}-%
\frac{1}{2(n-1)}r^{n-1}R\Big)+f_{R},_{v}r^{n-2}(r,_{u} e^{\varphi}),_{u}\Big].
\label{QJ}
  \ena
   One can check that the conserved charge $Q_J$ is just $M_{ms}$ in (\ref{MS}). We reach the same result by different routes. Therefore, (\ref{QJ}) and (\ref{MS}) should be a reasonable generalization of Misner-Sharp mass in the $n$-dimensional $f(R)$ gravity. At least, it inherits two significant properties of Misner-Sharp mass in Einstein gravity: it satisfies the unified first law and corresponds to a conserved charge associating to the Kodama vector. For $n=4$ and $k=1$ our result reduces the previous result in \cite{cai}.

   To display more deeper relation between the two routes, we prove the equivalence of the constraints (\ref{AvBu}) and (\ref{con}). We clearly calculate the first constraint (\ref{AvBu}),
   \be
   B_{,u}-A_{,v}=\frac{(n-2)V_{n-2}^{k}r^{n-2}e^{\varphi}}{16\pi G }\left[(\varphi,_{u}r,_{u}+
 r,_{uu})(f_{R},_{vv}+\varphi,_{v}f_{R},_{v})
  -(\varphi,_{v}r,_{v}+r,_{vv})(f_{R},_{uu}+\varphi,_{u}f_{R},_{u})\right].
   \en
  On the other hand, the LHS of (\ref{con}) yields,
   \be
   \nabla _{\mu }\nabla _{\nu}(f_{R}\nabla ^{\mu }K^{\nu})=e^{2\varphi}\left[(\varphi,_{u}r,_{u}+
 r,_{uu})(f_{R},_{vv}+\varphi,_{v}f_{R},_{v})
  -(\varphi,_{v}r,_{v}+r,_{vv})(f_{R},_{uu}+\varphi,_{u}f_{R},_{u})\right].
   \en
   It is clear that both of the two constraints are equivalent to (\ref{condi}).

  \section{static case}
  The spacetime (${\cal M},~ g$) becomes a stationary one if there exists a time-like Killing vector. Moreover, if it permits an $n-2$-dimensional maximally symmetric submanifold (${\cal K},~\gamma$) it is a static one, since the metric can be written in a time orthogonal coordinates system,
  \be
  ds^2=-l(r)dt^2+\frac{1}{h(r)}dr^2+r^{2}\gamma _{ij}dz^{i}dz^{j}.
  \label{metrics}
  \en
  In principle, for any concrete metric form one can always make coordinates transformations to write it in double null form (\ref{metricn}). And thus one can use all the results in the last section. Then one makes inverse coordinates transformation to get the results in the original coordinates. For the special status of static spacetime, we also explicitly present the Misner-Sharp mass in the static coordinates.

  To obtain the Misner-Sharp mass, we only need $T_{tt},~T_{rr},$ and $~T_{rt}$. The components of the field equation in this coordinate system read,
  \bea
  8\pi G T_{tt}&=&\frac{1}{2}fl+\frac{h}{r}\left[\left(\frac{n}{2}-1\right)f_Rl'-(n-2)lf_R'\right]+\frac{1}{4}f_R\left(h'l'-\frac{hl'^2}{l}+2hl''\right)-\frac{1}{2}f_R'lh'-hlf''_R,             \\
  8\pi G T_{rr}&=& -\frac{f}{2h}+\frac{1}{r}\left[(n-2)f'_R-\left(\frac{n}{2}-1\right)\frac{f_Rh'}{h}\right]+\frac{f_R}{4l}\left(\frac{l'^2}{l}-\frac{h'l'}{h}-2l''\right)+\frac{f_R'l'}{2l},
  \\
         T_{tr}&=&0,
         \ena
   where a prime denotes the derivative with respect to $r$. Again, the sectional curvature of $\cal K$ does not appear since $\cal M=K\times T$. The Ricci scalar reads,
   \be
   R=\frac{(n-3)(n-2)}{r^2}(k-h)-\frac{n-2}{r}\left(h'+\frac{hl'}{l}\right)+\frac{1}{2l}\left(\frac{hl'^2}{l}-h'l'-2hl''\right).
   \en
   In this case, the definition of Misner-Sharp mass (\ref{F}) becomes
   \be
 F\equiv dM_{ms}=A(t,r)dt+B(t,r)dr,
 \label{F1}
 \en
 where,
 \bea
A(t,r) &=&V_{n-2}^{k}r^{n-2}h(r,_{r}T_{tr}-r,_{t}T_{rr}), \\
B(t,r)
&=&V_{n-2}^{k}r^{n-2}\frac{(r,_{r}T_{tt}-r,_{t}T_{tr})}{l}.
\label{ABs}
\ena
 One immediately sees that $F$ in (\ref{F1}) is a closed form, since
 \be
 A_{,r}=B_{,t}=0.
 \en
 Hence, the Misner-Sharp mass is naturally well-defined for a static spacetime in the $n$-dimensional $F(R)$ gravity.

 Now we check the constraint in conserved charge method. The Kodama vector reduces to a Killing one in a static spacetime. In the static coordinates (\ref{metrics}), the time-like Killing vector reads,
  \be
  K=\sqrt{\frac{h(r)}{l(r)}}\frac{\partial}{\partial t}.
  \en
  It is easy to show that the constraint (\ref{con}) is also satisfied automatically. The Misner-Sharp mass, or the conserved charge associated to $K$ reads,
 \bea
  M_{ms}&=&\int_{\Sigma}Bdr\\
        &=&    \frac{V_{n-2}^{k}r^{n-3}}{8\pi G }\Big[\frac{n-2}{2}(k-I^{ab}\partial _{a}r\partial_{b}r)f_{R}+\frac{1}{2(n-1)}%
r^{2}(f-f_{R}R)- rI^{ab}\partial _{a}f_{R}\partial _{b}r\Big]~\notag \\
&+&\frac{V_{n-2}^{k}}{16\pi G }\int dr\Big[r^{n-2}h'+(n-2)r^{n-3}(h-k)+\frac{r^{n-1}}{n-1}R\Big]f_{R,r},
\label{MSs}
  \ena
  where $\Sigma$ marks a slice with constant $t$. For Einstein gravity with a cosmological constant, (\ref{MSs}) degenerates to
  \be
   M_{ms}=\frac{V_{n-2}^{k}r^{n-3}}{8\pi G }\left[\frac{n-2}{2}(k-I^{ab}\partial _{a}r\partial
_{b}r)f_{R}+\frac{r^2}{2(n-1)}(f-f_{R}R)\right].
  \en

   In a general $f(R)$ theory, the uniqueness theorem of the spherically symmetric space becomes invalid. $R$ may be not vanish even for  vacuum solutions. In that case the integration term can not be omitted.   Let's consider the 4 dimensional action in the following form,

   \be
   S=\frac{1}{16\pi G }\left(\int_{\cal M} d^{4}x\sqrt{-{\rm det}(g)}~R^{d+1}+\int_{\cal \partial M} d^{3}x\sqrt{-{\rm det}(h)}~2B_o\right),
  \label{action1}
  \en
   where $d$ is a constant. The corresponding field equation reads,
   \be
   (1+d)R^{d}R_{\mu \nu }-\frac{1}{2}R^{d+1}g_{\mu \nu }-(1+d)\nabla _{\mu }\nabla
_{\nu }R^{d}+g_{\mu \nu }(1+d)\square R^{d}=0.
\label{field}
  \en
  It is easy to see that the \sch~metric is a solution for the above equation. Obviously, the Misner-Sharp mass for the \sch~metric in $f(R)$ gravity is the same as in Einstein gravity, since \sch~ metric has a vanishing Ricci scalar $R$. The $f(R)$ gravity permits a non-trivial spherically symmetric solution other than the \sch~metric. The following metric also solves the field equation \cite{CB},
  \be
ds^{2}=-l(r)dt^{2}+\frac{dr^{2}}{h(r)}+r^{2}(d\theta ^{2}+\sin ^{2}\theta
d\phi ^{2})
\label{ansatzs}
\en
where
\bea
l(r)& =&r^{2d \frac{(1+2d )}{(1-d )}}+\frac{c}{r^{\frac{%
(1-4d )}{(1-d )}}}~, \\
h(r)& =&\frac{(1-d )^{2}}{(1-2d +4d ^{2})(1-2d (1+d
))}\left[ 1+\frac{c}{r^{\frac{(1-2d +4d ^{2})}{(1-d )}}}%
\right].
\ena
 Here $c$ is a constant, which reduces to the ~\sch~ mass parameter $c=-2M$ in Einstein gravity. This solution is called Clifton-Barrow solution. Now we check the physical sense of this solution by calculating its Misner-Sharp mass.
 With this metric the first part (none-integration part) in (\ref{MSs}) becomes
 \be
 \frac{6^d d (1+d) (2 d-1) \left(\frac{d (1+d)}{\left(2 d-1+2 d^2\right) r^2}\right)^d r^{\frac{1+d}{1-d}} \left(\frac{2 d \left(2 d-1+2 d^2\right) r^{\frac{2 d}{-1+d}}}{1-2 d}+\frac{c (-1+d)^2 r^{\frac{1+4 d^2}{d-1}}}{d}\right)}{2(4 d-1-6 d^2+4 d^3+8 d^4)},
 \en
 and the integration part reads,
 \bea
 \frac{6^d (1+d) \left(\frac{d (1+d)}{\left(2 d-1+2 d^2\right) r^2}\right)^d r^{\frac{1+d}{1-d}} \left(4 d^4 r^{\frac{2 d}{d-1}}+c r^{\frac{1+4 d^2}{d-1}}-4 c d r^{\frac{1+4 d^2}{d-1}}+d^3 \left(4 r^{\frac{2 d}{d-1}}-2 c r^{\frac{1+4 d^2}{d-1}}\right)+d^2 \left(5 c r^{\frac{1+4 d^2}{d-1}}-2 r^{\frac{2 d}{d-1}}\right)\right)}{2(4 d-1-6 d^2+4 d^3+8 d^4)} \notag\\
 +c_1,~~~~~~~~~~~~~~~~~~~~~~~~~~~~~~~~~~~~~~~~~~~~~~~~~~~~~~~~~~~~~~~~~~~~~~~~~~~~~~~~~~~~~~~~~~~~~~~~~~~~~~~~~~~~~~~~~~~~~~~~~~~~~~~~~~~~~~~~~~~~~~~~~~
 \ena
 where $c_1$ is an integration constant. The condition that (\ref{MSs}) reduces to the Misner-Sharp mass in Einstein gravity requires that
 \be
 c_1=-\frac{c}{2}.
 \en
 Finally, we reach the Misner-Sharp mass for the Clifton-Barrow solution,
 \be
 2GM_{ms}=(-c)^{(1-d)/(1-2d +4d ^{2})},
 \en
 which is independent on $r$, just like the \sch ~case.

   \section{conclusion}
   The unified first law is a significant approach in gravi-thermodynamics. Different from traditional black hole thermodynamics, in which the quantities depend on the asymptomatic behaviour of the whole manifold,  it concentrates on a finite region of the spacetime. Since there is no well-defined energy density of gravity field, the quasilocal mass plays critical role in unified first law. The Misner-Sharp mass is the proper mass form which guarantees the success of the unified first law. Moreover, we can derive the \sch~solution from Misner-Sharp mass form , in which we need not solve the field equation. It deserves to explore the generalization of Minser-Sharp mass in an $n$-dimensional modified gravity. $f(R)$ gravity has the essential properties of higher order derivative theory, as well as high operability. In this article we obtain the Misner-Sharp mass for $f(R)$ gravity in the $n$-dimensional spacetime with $n-2$-dimensional maximally symmetric submanifold by using the ``inverse first law" method in a double-null coordinate system. In this method, we first assume the unified first law still takes the same form as in Einstein gravity. And then we identify the required mass form as the generalized Misner-Sharp mass in this case. We find a constraint for a well-defined Misner-Sharp mass. Also, we derive the Misner-Sharp mass by the conserved charge method. This method present an exactly the same mass form and an equivalent constraint for the existence of a well-defined Misner-Sharp mass. So, it seems a reasonable generalization of the Misner-Sharp mass in the $n$-dimensional $f(R)$ gravity.

   Static spacetime is an important special case in gravity theory. We also present the implicit form of the Misner-Sharp mass in a static spacetime in the $n$-dimensional $f(R)$ gravity. Cifton-Barrow solution is an interesting solution in $f(R)$ theory. As an example, we calculate the Misner-Sharp mass for this solution. We find that it is a constant which is independent on the radius $r$, just as the situation of the \sch~solution.

 {\bf Acknowledgments.}
   This work is supported by the Program for Professor of Special Appointment (Eastern Scholar) at Shanghai Institutions of Higher Learning, National Education Foundation of China under grant No. 200931271104, and National Natural Science Foundation of China under Grant Nos. 11075106 and 11275128.

\end{document}